\documentstyle[aps,multicol,epsfig,pre]{revtex}

\begin{document}
\draft
\preprint{HEP/123-qed}
\title{Time dependence of breakdown in a global fiber-bundle model
with continuous damage}
\author{L. Moral$^1$, Y. Moreno$^2$, J.B.G\'{o}mez$^1$, A. F. Pacheco$^1$}
\address{
$^1$ Facultad de Ciencias, Universidad de Zaragoza,
50009 Zaragoza, Spain.\\ $^2$ The Abdus Salam International Centre for
Theoretical Physics,\\ Condensed Matter Group, P.O. Box 586, Trieste,
I-34014, Italy.}
\date{\today}
\maketitle
\widetext
\begin{abstract}
A time-dependent global fiber-bundle model of fracture with
continuous damage is formulated in terms of a set of coupled
non-linear differential equations. A first integral of this set is
analytically obtained. The time evolution of the system is studied
by applying a discrete probabilistic method. Several results are
discussed emphasizing their differences with the standard
time-dependent model. The results obtained show that with this
simple model a variety of experimental observations can be
qualitatively reproduced.
\end{abstract}
\begin{multicols}{2}
\narrowtext


Fracture in disordered media has attracted much scientific and
industrial interest for many years
\cite{h90,cha,garcia97,maes98,petri94,zape97,prlus00}; it is,
however, a complex problem for which a definite physical and
theoretical treatment is still lacking. An important class of
models of material failure is the fiber-bundle models (FBM) which
have been extensively studied during the past decades
\cite{prlus00,col57,pho83,new94,new95,vaz99}. These models consist
of a set of parallel fibers having statistically distributed
strength. The sample is loaded parallel to the fiber direction,
and a fiber fails if the load acting on it exceeds a threshold
value. When a fiber fails, its load is transferred to other
surviving fibers in the bundle, according to a specific transfer
rule. Among the possible options of load transfer, one
simplification that makes the problem analytically tractable is
the assumption of equal load sharing (ELS), or global load
transfer, which means that after each fiber breaks, its stress is
equally distributed among the intact fibers. Thus, the ELS option
constitutes a sort of mean field approximation to other more
realistic rules of stress transfer where a stress enhancement
occurs in the neighborhood of failed elements. So far, the failure
rule applied in standard FBM is discontinuous and irreversible,
i.e., when the local load exceeds the failure threshold of a
fiber, the fiber is removed from the calculation and is never
restored. Recently, a novel continuous damage law was introduced
in these models \cite{nat97,kun00}. Thus, when the strength
threshold of a fiber is exceeded, it yields, and the elastic
modulus of the fiber is reduced by a factor $a$ ($0<a<1$).
Multiple yields of a given fiber are allowed. It is argued that
this description of damage in terms of a continuous parameter
corresponds to the consideration of the system at a length scale
larger than the typical crack size; i.e., if the smallest elements
of the model are the fibers, the continuous damage is due to
cracking inside the fibers. This generalization of the standard
FBM is suitable to describe a variety of elasto-plastic
constitutive behaviors \cite{na95,evans94,kanada99}.

FBM come in two settings, static and time-dependent or dynamic.
The static version of FBM simulates the failure of materials by
quasiestatic loading, i.e, by a steady increase in the load over
the system up to its macroscopic failure. The stress on each fiber
is the independent variable and the strength of each element is
the distributed random variable. On the other hand, the dynamic
FBM simulates failure by creep rupture, static fatigue, or delayed
rupture, i.e., a (usually) constant load is imposed on the system
and the elements break because of fatigue after a period of time.
The time elapsed until the system collapses is the lifetime of the
bundle. Time acts as an independent variable, and the initial
lifetime of each element, for a prescribed initial stress, is the
independent identically distributed random quantity.

The concept of continuous damage in FBM has only been applied to
the static setting. However, time-dependent mechanisms also play a
key role in the process of fracture. Phenomena such as fatigue
and stress corrosion are of utmost importance for real
applications. These time-dependent effects have not been included
before in continuous damage descriptions. So, the precise purpose
of this paper is to formulate for the first time, the
time-dependent FBM with continuous damage and compute the
differences appearing with respect to the standard dynamic FBM.

In these models the most widely used breaking rate function is the
power law \cite{vaz99}, in which elements break at a rate
proportional to a power of their stress, $\sigma^{\rho}$, where
the exponent $\rho$ is an integer called the stress corrosion
exponent. This type of breaking rate will be assumed here.

Our analysis will be restricted to the global transfer modality,
and we will assume that the size of the bundle, $N$, is very
large. This enables us to formulate the evolution of the system in
terms of continuous differential equations. This type of
equations, similar to those appearing in radioactivity, was first
used by Coleman \cite{col57}, and later in \cite{new95}. At this
point, it is worth recalling that for the standard model, the
lifetime $T$ of the bundle can be analytically obtained. In this
case, the differential equation governing the time evolution of
the system reads
\begin{equation}
\frac{dN_0}{dt}=-N_0 \cdot f^{\rho},
\label{eq1}
\end{equation}
where $f=\sigma=N/N_0$ is the strain of the bundle assuming that the
elastic modulus of the fibers is $Y=1$ and $\sigma$ is the individual stress
acting on one fiber. The solution of Eq.\ (\ref{eq1}) should fulfill
the condition $N_0(t=0)=N$. The integration of\ (\ref{eq1}) is
straightforward and the lifetime of the bundle is given by $T=\frac{1}{\rho}$.

Now, suppose an ELS bundle formed by $N$ fibers which breaks
because of stress corrosion under the action of an external
constant load $F=N\cdot \sigma_0$, with $\sigma_0=1$. The breaking
rate of the fibers, $\Gamma$, is assumed to be of the power-law
type, $\Gamma=\sigma^{\rho}$. As before, $f$ will denote the
strain of the bundle and $Y=1$ will represent the initial
stiffness of the individual fibers. The original dynamic FBM is
generalized by allowing that one fiber can fail more than once,
and thus we define the integer $n$ as the maximum value of
failures allowed per fiber. Besides, the parameter $a$ ($<1$) will
represent the factor of reduction in the stiffness of the fibers
when they fail. As up to $n$ partial yielding events are permitted
per fiber, at any time the population of fibers will be sorted in
$n+2$ lists. Thus $ N=N_0+N_1+\dots+N_n+N^{\prime}, \label{eq3} $
where $N_j$ ($j=0,\dots,n$) denotes the number of elements that
have failed $j$ times. $N^{\prime}$ denotes the number of elements
that have failed $n+1$ times and therefore are inactive (i.e.,
they do not support load anymore). At $t=0$, the $N$ elements of
the bundle form the list $0$, $N_0=N$, and at $t=T$,
$N^{\prime}=N$. The specification, at a given time $t$, of the
value of $N_j$, for $j=0,1,\dots,n$, provides the state of the
system. In our continuous formulation the $N_j$ will be real
positive numbers lower than $N$.

As the external load $F=N$ is supported by the present active fibers,
we have
\begin{equation}
N=f\cdot(N_0+a N_1+a^{2}N_2+\dots+a^{n}N_n),
\label{eq4}
\end{equation}
and hence $f=N/(N_0+aN_1+a^{2}N_2+\dots+a^{n}N_n)$.

The time evolution equations are:
\begin{eqnarray}
 \frac{dN_0}{dt}=f^{\rho}(-N_0), \nonumber \\
 \frac{dN_1}{dt}=f^{\rho}(N_0-kN_1), \nonumber \\
 \frac{dN_2}{dt}=f^{\rho}k(N_1-kN_2), \label{eq5} \\
 \vdots \qquad \qquad \qquad\qquad \vdots \nonumber \\
 \frac{dN_n}{dt}=f^{\rho}k^{n-1}(N_{n-1}-kN_n), \nonumber
\end{eqnarray}
where the ubiquitous constant factor $k$ represents $k=a^{\rho}$.
This is a system of coupled, first-order, non-linear differential
equations. Its solution must fulfill the initial condition
\begin{eqnarray}
N_0(t=0)=N \nonumber \\ N_j(t=0)=0, \quad j\ne 0. \label{eq6}
\end{eqnarray}

On the right-hand side of Eq.\ (\ref{eq5}), the positive terms
represent the sources and the negative ones represent the sinks of
the various lists. Eq.\ (\ref{eq5}) does not have an analytic
solution. However a first integral can be given expressing $N_j$
($j\ne 0$) in terms of $N_0$. The source of non-linearity in Eq.\
(\ref{eq5}) is the factor $f^{\rho}$ on the right-hand side. This
factor can be eliminated by reformulating the system of equations
in such a way that $N_0$ is the new independent variable and $N_j$
($j\ne 0$) the dependent variables. Denoting by a prime the
derivative with respect to $N_0$, one easily obtains

\begin{eqnarray}
 N_1^{\prime}=(kN_1-N_0)/N_0, \nonumber \\
 N_2^{\prime}=k(kN_2-N_1)/N_0, \label{eq7} \\
 \vdots \qquad \qquad \qquad\qquad \vdots \nonumber \\
 N_n^{\prime}=k^{n-1}(kN_n-N_{n-1})/N_0. \nonumber
\end{eqnarray}

This coupled set of first-order linear differential equations must
fulfill
\begin{equation}
N_j(N_0=N)=0, \quad j\ne 0 \label{eq8}
\end{equation}
as initial conditions.

The solutions of Eq.\
(\ref{eq7}) fulfilling Eq.\ (\ref{eq8}) for an arbitrary index $l$,
$l=1,2,\dots,n$, are
\begin{equation}
N_l=\sum\limits_{j=0}^{l}a_j^{(l)}N_0^{k^{j}}.
\label{eq10}
\end{equation}

This ansatz is easily proved by induction, and the $a_j^{(l)}$
coefficients are recursively calculated,
\begin{eqnarray}
 a_j^{(l+1)}=\frac{k^{l}a_j^{(l)}}{(k^{l+1}-k^{l})},\quad j=0,1,2,\dots,l
 \nonumber \\
 \qquad \label{eq11} \\
 a_{l+1}^{(l+1)}=\frac{\left(-\sum\limits_{j=0}^{l}a_j^{(l+1)}N^{k^{j}}\right)}{N^{k^{l+1}}}.\qquad
 \nonumber
\end{eqnarray}
These exact functions $N_j=N_j(N_0)$, $j\ne 0$ can be used as the
base of an elegant numerical method suited to compute the
time-dependent solution of Eq.\ (\ref{eq5}). This will be
commented on in a forthcoming publication. In this paper, however,
we will use Eq.\ (\ref{eq10}) merely to test the accuracy of
another approximated method, a discrete probabilistic one
\cite{vaz99,curtin91,gomez98}, that will be used to solve Eq.\
(\ref{eq5}). In this case of an ELS model with continuous damage,
the elementary time step for one fiber to yield is given by
\cite{vaz99}:
\begin{equation}
\delta=\frac{1}{N_0f^{\rho}+
N_1(af)^{\rho}+N_2(a^2f)^{\rho}+\dots+N_n(a^nf)^{\rho}},
\label{eq12}
\end{equation}
with
\begin{equation}
f=\frac{N}{N_0+aN_1+a^2N_2+\dots+a^nN_n}. \label{eq13}
\end{equation}
Thus, $\delta$ is the inverse of the total ``decay width'' of the
system. The total decay width is the sum of the contribution of
all the lists $j=0,1,\dots,n$. And each list contributes with a
term
\begin{equation}
\Gamma_j=N_j\sigma_j^{\rho}=N_j(a^jf)^{\rho}.
\label{eq14}
\end{equation}

The probability that the individual failure takes place in the
list $j$ is equal to $p_j=\Gamma_j\cdot \delta$. With this natural
assignment of probabilities, it is apparent that
$\sum\nolimits_{j=0}^n p_j=1$. In the process of breaking, a set
of $N$ elements and $n$ allowed partial yields, we will have a
total of $N(n+1)$ deltas, whose sum $\sum\nolimits_{i=1}^{N(n+1)}
\delta_i=T$, is the lifetime of the bundle. This method starts
from an initial state,
$(N_0,N_1,N_2,\dots,N_n,N^\prime)=(N,0,0,\dots,0,0)$ and ends in
$(0,0,0,\dots,0,N)$.

\begin{figure}[b]
\begin{center}
\epsfig{file=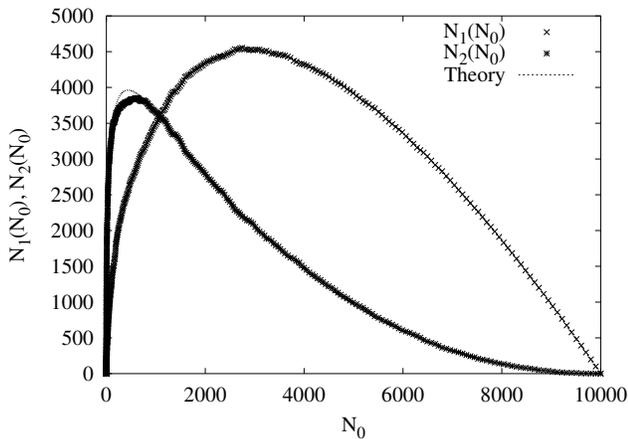,width=8.5cm,angle=0,clip=1}
\end{center}
\caption{Results of the probabilistic method (dotted curves) compared
with the analytical prediction Eq.\ (\ref{eq10}). The parameters used in the simulation are $N=10^4$, $a=0.8$, and
$\rho=2$. The fibers are allowed to break $n=4$ times.}
\label{figure1}
\end{figure}

\begin{figure}[b]
\begin{center}
\epsfig{file=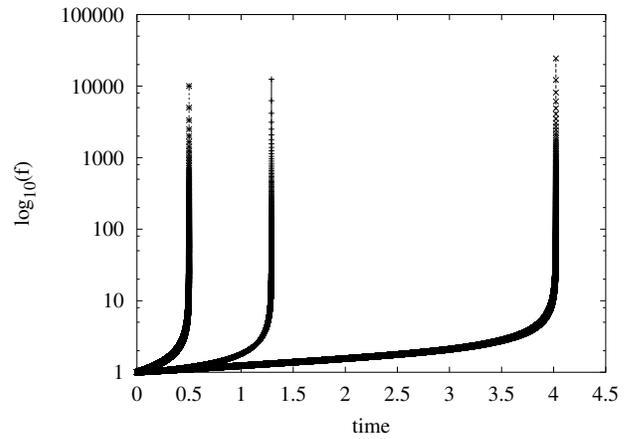,width=8.5cm,angle=0,clip=1}
\end{center}
\caption{Bundle's strain ($f$) as a function of dimensionless time
for different values of $n$ (from left to right $n=0,1,4$). The
model parameters are the same as Fig\ \ref{figure1}. Multiple
failures lead to increase in the lifetime of the bundle.}
\label{figure2}
\end{figure}

\begin{figure}[b]
\begin{center}
\epsfig{file=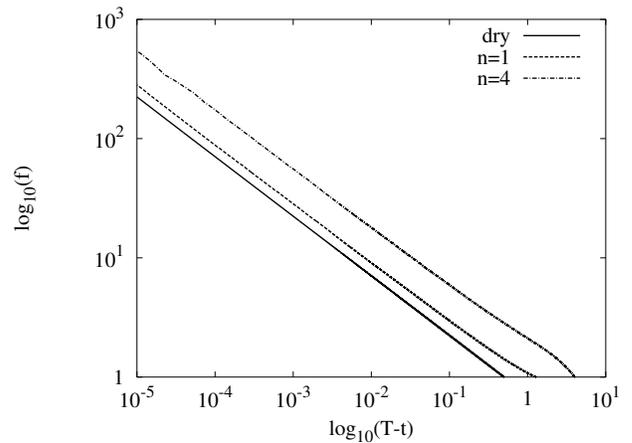,width=8.5cm,angle=0,clip=1}
\end{center}
\caption{Behavior of the material strain near the macroscopic point of
rupture for $N=10^4$, $a=0.8$, $\rho=2$ and the values of $n$
indicated in the figure. The scaling relation satisfies $f\sim
(T-t)^{\beta}$, with $\beta=-\frac{1}{\rho}$.}
\label{figure3}
\end{figure}

The results obtained with this method are shown in Figs.\
\ref{figure1}-\ref{figure3}.  In each simulation, one has to fix
the vector $(N,\rho,a,n)$, under the hypothesis that $F=N$ and
$Y=1$. The standard model is obtained by setting $n=0$. In Fig.\
\ref{figure1}, for a vector $(10^{4},2,0.8,4)$, we compare the
prediction of this method for $N_1(N_0)$ and $N_2(N_0)$ with the
analytic result expressed in equation\ (\ref{eq10}). It is
apparent that the agreement is excellent. In Fig.\ \ref{figure2},
the behavior of the bundle strain, $f$, as a function of time, is
plotted for the same vector of parameters as before but
considering also the cases of $n=0$ and $n=1$. Here, one
appreciates how the hypothesis of continuous damage leads to
increasing the bundle's lifetime. It is interesting to note that
at the time at which the system collapses, the relation
$F=a^{n}f_c$ holds, {\em i.e.}, the ultimate strain of the bundle,
$f_c$, is related to the external load imposed over the system
through the factor $a^{-n}$. This is also the case for the static
continuous damage model \cite{kun00} but substituting $F$ by
$F/N$. Finally, Fig.\ \ref{figure3} shows the behavior of the
bundle's strain near the point of macroscopic rupture. It turns
out that the strain scales with the distance to the critical
point, $T-t$, as $f\sim (T-t)^{\beta}$ with
$\beta=-\frac{1}{\rho}$. This result is universal, {\em i.e.}, the
material approaches its macroscopic collapse in this way
regardless of the number of allowed partial yields.  Universal
behaviors are quite important in fracture processes because of the
intrinsic and unavoidable sample to sample variations. Thus,
although the lifetime of the bundle varies from one realization to
another, the strain of the bundle would satisfy, for a given
$\rho$ value, the same scaling function near the point of
macroscopic collapse regardless of $n$. It is worth noting that
this feature is not observed only in our simple model. For
example, in a similar (in spirit) although more complicated model
where visco-elastic cells are introduced, the same behavior near
the critical point is found \cite{kunpc}.

Up to now, we have assumed that the external load acting on the
bundle is $F=N$. If we consider that $F$ is modified in a factor
$\phi$, $F_{\phi}=\phi N$, then the lifetime of reference $T$
would be modified in the following way,
\begin{equation}
T_{\phi}=\frac{T}{\phi^{\rho}}. \label{eq16}
\end{equation}
Thus, we see that this exact property that works for the standard
dynamic FBM with any load transfer rule and the power-law breaking
rate \cite{pho83} extends to the continuous damage version of this
model.

Throughout this paper, we have considered brittle failure after
$n$ damage events. Nevertheless, in order to describe macroscopic
strain hardening instead of global failure, we should allow the
fibers to have a $a^{n}$ residual stiffness after having yielded
$n$ times.

\begin{figure}[b]
\begin{center}
\epsfig{file=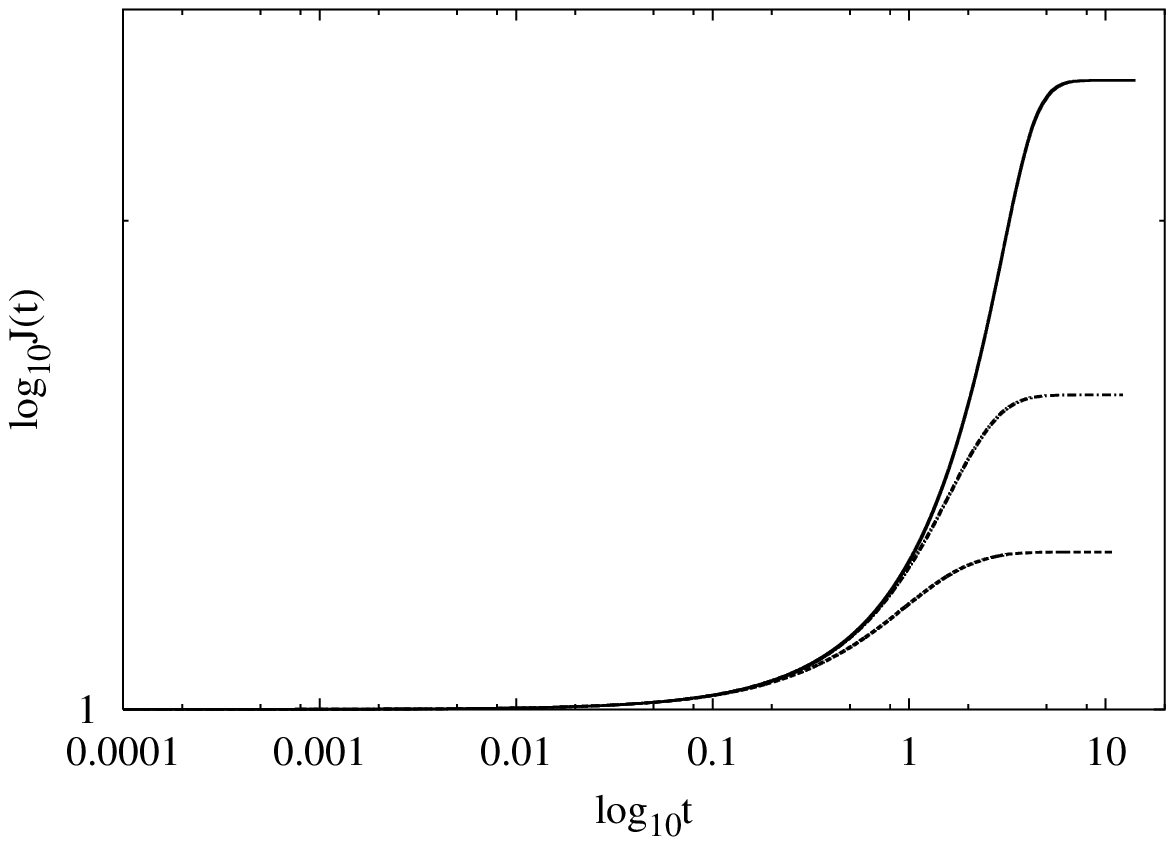,width=8.5cm,angle=0,clip=1}
\end{center}
\caption{The creep compliance $J(t)$ as a function of
dimensionless time for different values of $n$ ($n=1,2,4$ as $J$
increases). The behavior is qualitatively the same as that
obtained in creep experiments on amorphous materials.}
\label{figure4}
\end{figure}

Figure\ \ref{figure4} shows the creep compliance $J(t)$, defined
as $ J(t)=f(t)/F$, as a function of dimensionless time $t$ for
different values of $n$. It can be seen that the model
qualitatively reproduces the behavior of $J(t)$ observed in
experiments on amorphous materials \cite{plastic82}. For short
times, the compliance is very slowly time dependent. As time
passes, $J(t)$ becomes very strongly time dependent and finally
the system approaches a plastic state for which the compliance is
again very slowly time dependent. This confirms that the behavior
of $J(t)$ depends on the time scale of the experiment
\cite{plastic82}.

In short, we have introduced a novel time dependent model of
fracture with continuous damage for the breakdown of materials.
The model was formulated in terms of a set of coupled non-linear
differential equations and the time evolution of the system was
studied by applying a discrete probabilistic method. This model is
potentially useful to describe some elasto-plastic behaviors
observed in real material fracture processes. Besides, it could
guide our understanding to more complex time dependent models of
fracture with continuous damage.


A.F.P thanks Javier Ribera for discussions. Y.M thanks A. Vespignani,
H. J. Herrmann and F. Kun for stimulating discussions on our model,
and the Abdus Salam International Centre for Theoretical Physics for
hospitality and financial support. This work was supported in part by
the Spanish DGICYT Project PB98-1594.

\end{multicols}

\begin{references}
\bibitem{h90} {\sl Statistical
Models for the Fracture of Disordered Media\/}. Editors, H.J. Herrman and S. Roux, North Holland
(1990), and references therein.
\bibitem{cha} {\sl Statistical Physics of Fracture and Breakdown in
Disordered Systems\/}. B. K. Chakrabarti, L. G. Benguigui, Clarendon
Press, Oxford (1997), and references therein.
\bibitem{garcia97} A. Garcimartin, A. Guarino, L. Bellon,
S. Ciliberto, {\em Phys. Rev. Lett.} {\bf 79}, 3202 (1997).
\bibitem{maes98} C. Maes, A. van Moffaert, H. Frederix, H. Strauven,
{\em Phys. Rev. B} {\bf 57}, 4987 (1998).
\bibitem{petri94} A. Petri, G. Paparo, A. Vespignani, A. Alippi,
M. Constantini, {\em Phys. Rev. Lett.} {\bf 73}, 3423 (1994).
\bibitem{zape97} S. Zapperi, P. Ray, H. E. Stanley, A. Vespignani,
{\em Phys. Rev. Lett.} {\bf 78}, 1408 (1997).
\bibitem{prlus00} Y. Moreno, J. B. Gomez, A. F. Pacheco,{\em
Phys. Rev. Lett.} {\bf 85}, 2865 (2000).
\bibitem{col57} B.D. Coleman, {\em J. Appl. Phys.} {\bf 28}, 1058
(1957); {\it ibid} {\bf 28}, 1065 (1957).
\bibitem{pho83} S.L. Phoenix and L. Tierney, {\em Eng. Frature Mech.} {\bf 18}, 193 (1983).
\bibitem{new94}  W.I. Newman, A.M. Gabrielov, T.A. Durand,
S.L. Phoenix and D.L. Turcotte, {\em Physica D}{\bf 77}, 200 (1994).
\bibitem{new95} W.I. Newman, D.L. Turcotte and A.M. Gabrielov, {\em
Phys. Rev. E} {\bf 52}, 4827 (1995).
\bibitem{vaz99} M. V\'{a}zquez-Prada,J. B. G\'{o}mez, Y. Moreno, A. F.
 Pacheco, {\em Phys. Rev. E} {\bf 60}, 2581 (1999).
\bibitem{nat97} S. Zapperi, A. Vespignani, and H. E. Stanley, Nature
{\bf 388}, 658 (1997).
\bibitem{kun00} F. Kun, S. Zapperi, H. J. Herrmann, {\em
Eur. Phys. J.} {\bf B17}, 269 (2000).
\bibitem{na95} A. E. Naaman, H. W. Reinhardt, {\em High performance
fiber reinforced cement composites} (E and FN Spon, London, 1995).
\bibitem{evans94} A. G. Evans, J. M. Domergue, E. Vagaggini, {\em
J. Am. Cerramic Soc.} {\bf 77}, 1425 (1994).
\bibitem{kanada99} T. Kanada, V. C. Li, {\em J. Eng. Mech.} {\bf 125},
290 (1999).
\bibitem{curtin91} W.A. Curtin and H. Scher, {\em Phys. Rev.
Lett.} {\bf 67}, 2457 (1991).
\bibitem{gomez98} J.B. G\'omez, Y. Moreno, A. F. Pacheco, {\em
Phys. Rev. E.}  {\bf 58}, 1528 (1998)
\bibitem{kunpc} F. Kun, private communication.
\bibitem{plastic82} D. Froelich, in {\em Plastic deformation of amorphous and
semi-crystalline materials}, edited by B. Escaig and C. G'Sell, (Les
Editions de Physique, Les Ulis, France) 1982.

\end{references}
\end{document}